# Planar and non-planar solidification

# of optically organized smectic films

H. Vasseur

PSC - 33 rue St Leu, 80039 Amiens Cedex, France, EU

Freely suspending films of smectic liquid crystals exhibit a succession of phases providing a good mean for understanding almost two-dimensional phase transitions. In 1978 Young, Pindak, Clark and Meyer[1] have observed the behavior of very thin films theoretically predicted by G. Friedel in 1922[2]. The main tool used for studying the structure of liquid films consists of x-ray scattering experiments pioneered by Moncton, Pindak, Davey, and Brown[3]. More recently these studied have been extended to solid samples obtained by freezing symmetric multiphase suspended smectic films.

We present in this letter an original freezing process yielding remarkably homogeneous films of chiral smectics. This optical homogeneity is observed on planar films as well as on films exhibiting various complex three-dimensional shapes. The principle is very simple: After forming the films by spreading the liquid crystal above a hole in a glass slice placed over a hot stage, the film is heated from below. The hot film is exposed to ambient temperature. Then, a solid object at room temperature with a specifically adapted shape is immersed in the liquid film. The constraints imposed by the object curves the film and stabilized various solid 2D and 3D structures. These solids exhibit homogeneous optical properties due to long-range organization of the molecular orientation (tilt) together with a complex helical arrangements of the frozen smectic films.

**Planar solid films** (Fig.1-2):

A thin rectangular monocrystal of MgO (with long side L=1mm, short side S=0.5 mm) is partially immersed in the matrix of a thick film of cholesteryl myristate (SmA phase at T=88K ). The solid phase grows around the MgO section in few seconds (with about $2.10^{-5}$ m thickness). The solidification is provoked by the contact of the liquid film with both air and MgO. In a 2mm-diameter film, we obtain a 1mm-width crystalline crown. The geometry of the solidified crown reflects that of the MgO substrate: it is constituted of four homogeneous, two of them (that we denote by L) are located along the long side of the substrate, whereas the two parts denoted by S are along the short sides. L and S are connected by circular regions close to the corners of the substrate.

Optical observations show that L and S appear black under crossed analyzer-polarizer (A/P) when the analyzer is set parallel to the corresponding substrate sides. This indicates that the molecules are homogeneously oriented in each macroscopic part of the solidified film (Fig. 1). Let us note that we obtain such "monodomain" molecular ordering only when the MgO substrate is perfectly clived. In addition, their projection in the sample plane is either parallel or normal to the substrate side. Otherwise, we observe strong orientational fluctuations destroying the optical homogeneity of the samples. In the neighborhood of the MgO corners, the molecular orientation varies quasi circularly between their directions in S and L.

We have determined the precise molecular orientation (parallel or normal to the substrate sides) in the solidified film by tilting it around the analyzer axis (parallel to L) : S turns to white (Fig. 2). S doesn't turn in white by tilting the sample around the polarizer axis (parallel to S). That means that the molecular projections are normal to the substrate sides Moreover, this optical behavior demonstrates that the molecules are tilted within the smectic layers, otherwise the whitening of S monodomains could not be observed, in agreement with observations in solid polycrystalline myristate[4].

In a similar experiment, in which the substrate is replaced by a cylindrical needle[5], the solid crown is circular. Under crossed P/A it exhibits a black cross characteristic of radial or orthoradial molecular orientation. When tilting the solid sample, the two black arms of the cross parallel to the polarizer move towards one of the arms parallel to the analyzer, indicating an orthoradial organization. The arms shift angle permits to measure the tilt angle[6].

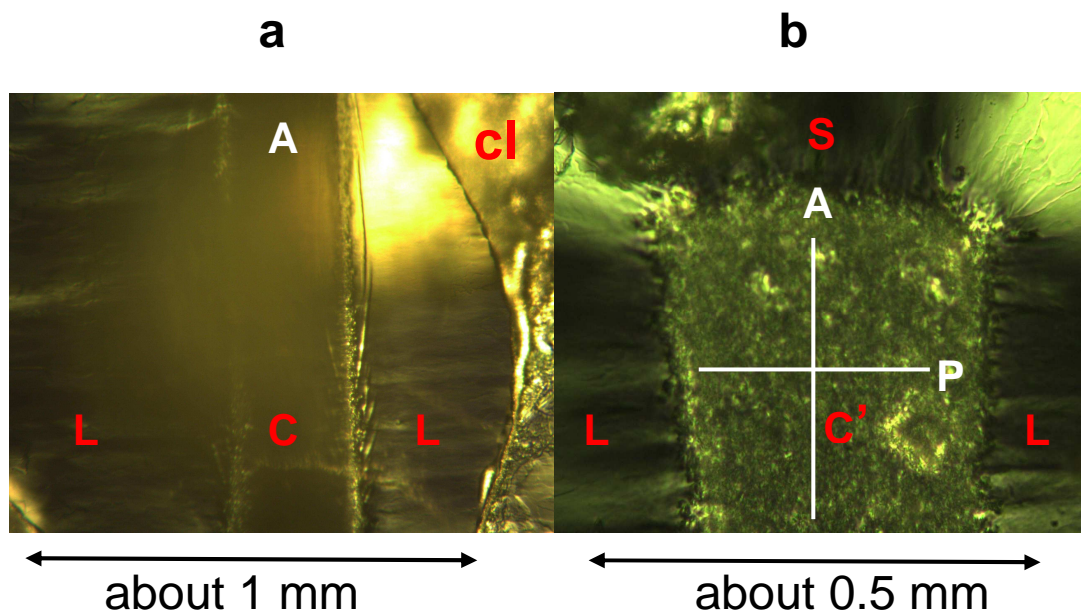

**a**  **b**

about 1 mm    about 0.5 mm

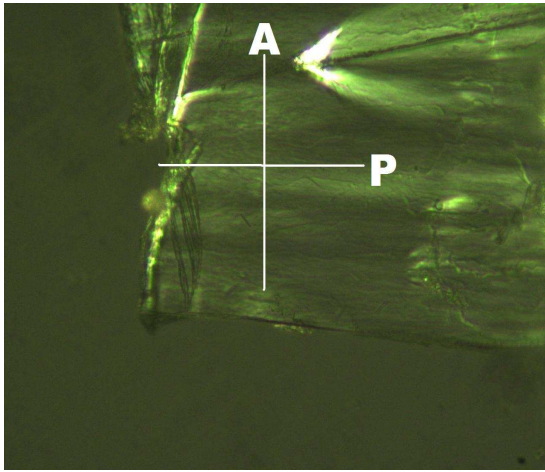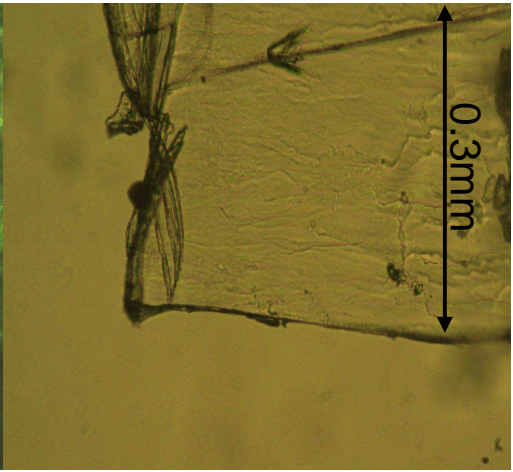

c  d

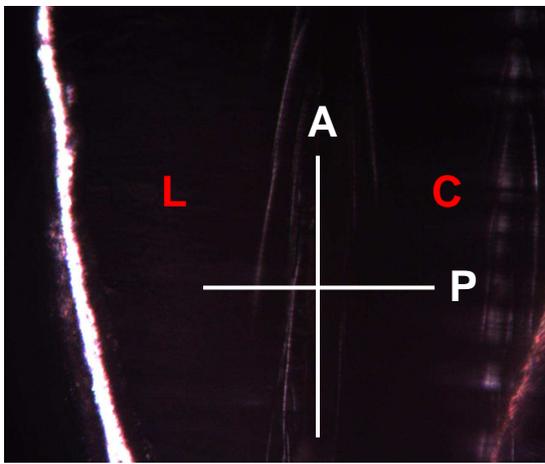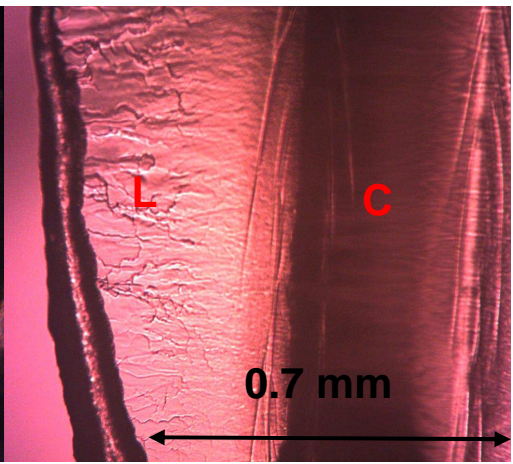

c'  d'

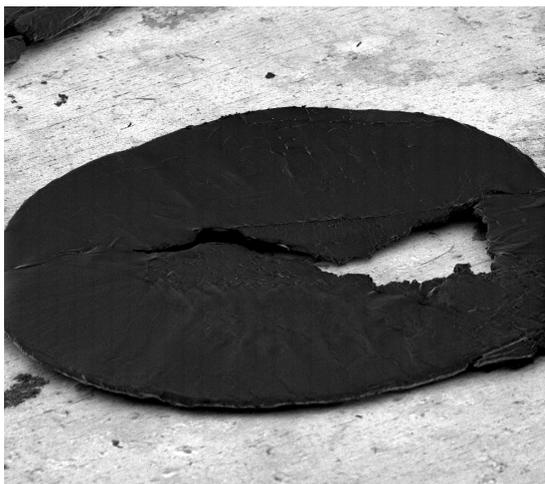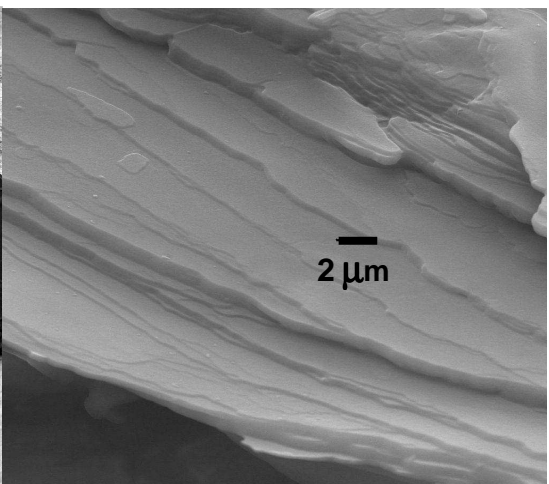

e  f

*Figure 1 :a) Liquid crystal (CL)-solid (S) interface during the crystal growth. C indicates the MgO substrate. The solidified part of the film appears black under crossed P/A because the molecules are parallel to P or A. b) The same film after solidification and extraction of the MgO crystal. The print (C') of the crystal on the film substrate is polycrystalline. L and S monodomains parallel to the sides of the substrate appear black. In the circular parts of the solid crown which neighbors the MgO corners the molecular orientation varies continuously yielding white color. c) Detail of a L monodomain under crossed P/A. The white spot indicates a defect in macroscopically homogeneous molecular orientation. D) The same detail under parallel P/A. c') A solidified film with a better molecular orientation under crossed P/A. d') The same solidified film under parallel P/A.  f,g) SEM images of the solidified film. The strata seen in Fig.f contain a few smectic layers*

Observations of a solidified film under electronic scanning microscope (SEM) reveals a layered structure. Each layer is composed of a few smectic layers. They curl to form a macroscopic elliptic helix around the MgO crystal. The systematic value of the helix handedness proves the chiral origin of this complex and intriguing smectic planes organization.

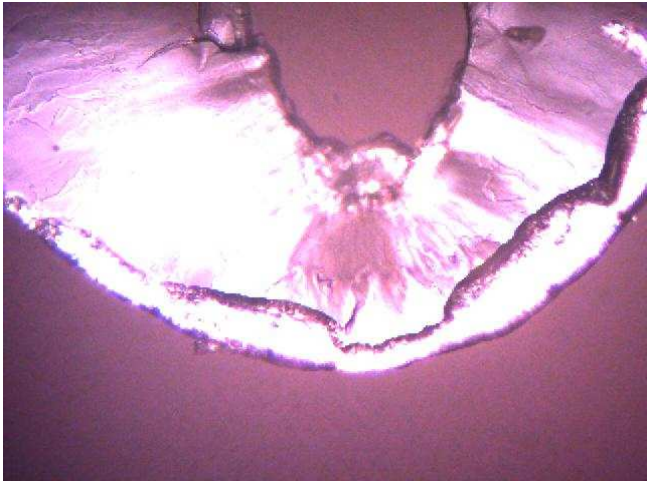

**a**

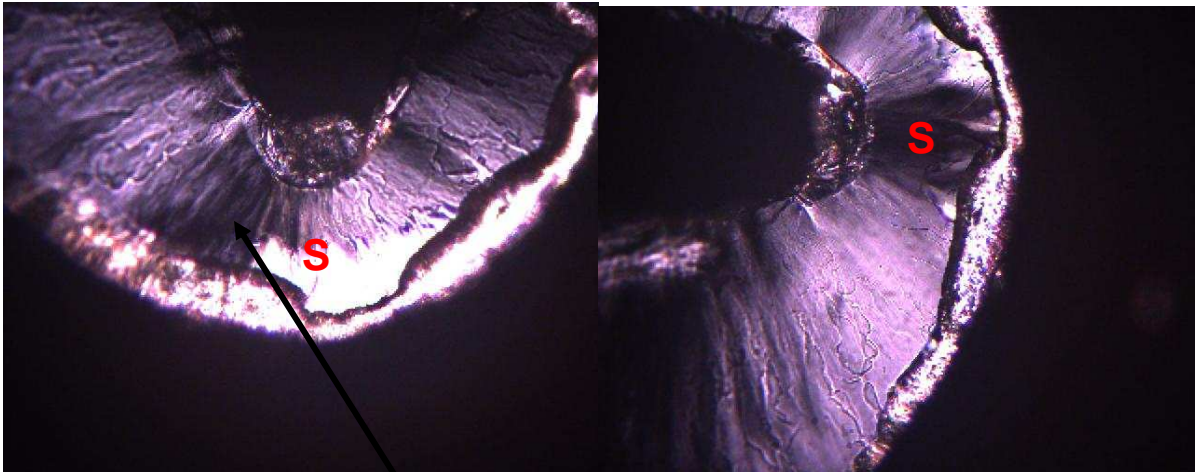

**b**                                    **c**

The black arm shifts in circular part

of the solidified film

*Fig. 2 : a) Solidified film no tilted with great contrast. The circular parts which join L and S appears white because the mean molecular tilt is not parallel to analyzer axis and the polarizer axis. b) By tilting (15°) the solidified film around analyzer axis (parallel to L). S turns to white and and the black arm shifts in the circular part. c) By tilting around polaryzer axis (parallel to S). S remains black. The L domain doesn't really turn in white, that means the*

*molecular tilt within layers is small and their average projection in the solid plane remains almost normal to L.*

**Non-planar films (Fig.3):**

For obtaining non-planar solid films we have replaced the crystalline substrate by a metal ring, superficially immersed at the surface of the smectic film. When the ring is slowly raised from the film surface, a solid empty cylinder of liquid crystal is formed between the ring and the film. The top of the cylinder is solidified. Progressively rising the ring increases the length of the cylinder, which bottom is feed by the molecules of the liquid film. The diameter of the cylinder sections can be modified during the formation process by modifying the raising speed. We obtain thus solids with various 3D shapes. SEM observations show they exhibit layered structures similar to that observed (Fig. 1f) in planar films. The layers wind around the cylinder axis (Fig. 3d). Under crossed P/A microscope, a black line parallel to the cylinder axis appears at its surface (Fig. 3a, c). When P is parallel to A, the cylinder appears transparent totally. Thus, the molecules are tilted within the smectic layers, and the projection of the mean molecular orientation is parallel to the cylinder axis. The molecular orientation structure has a perfectly cylindrical organization.

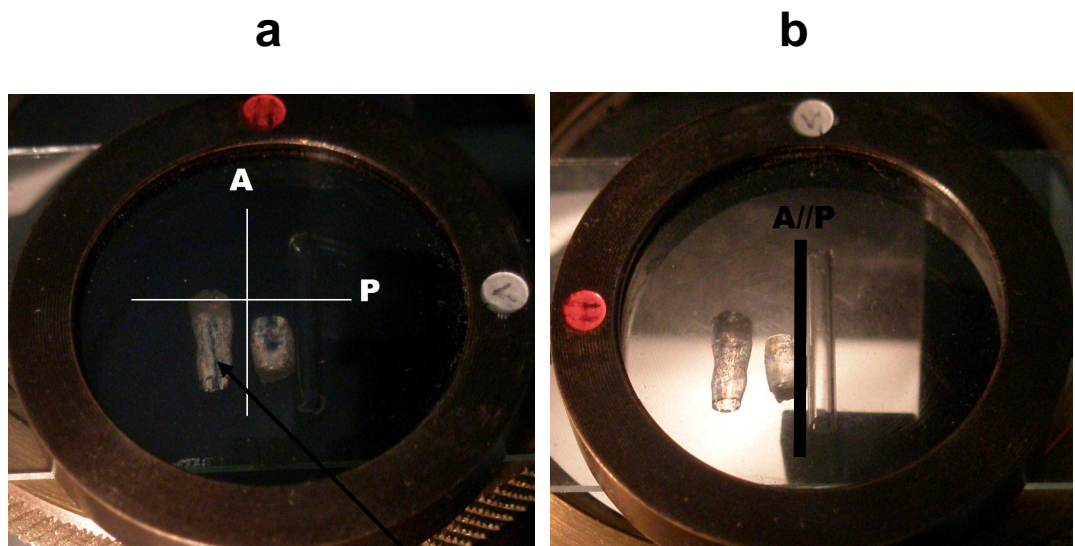

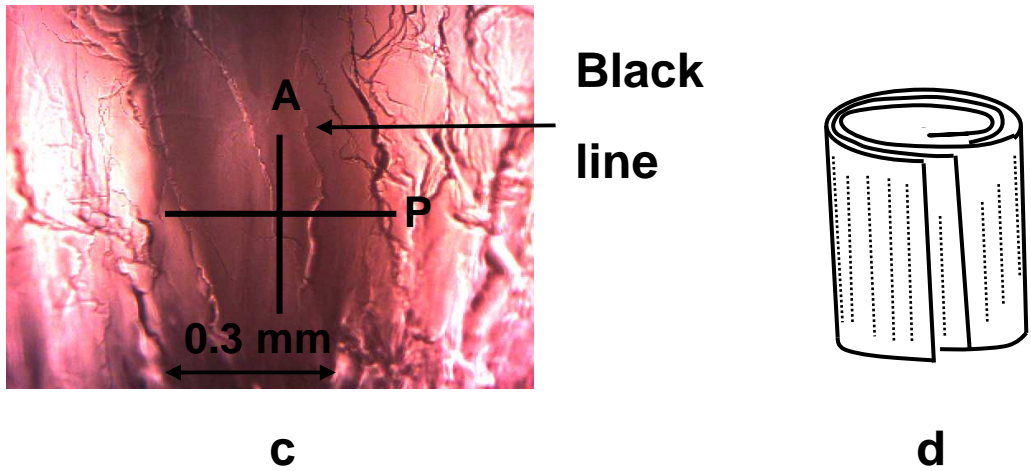

*Figure 2: a) Cylinders of cholesteryl myristate observed under analyzer and polarizer crossed. A black line characteristic of the homogenous molecular distribution appears on the two solidified cylinders. On the right, a reference glass pipe which appears black. b) Cylinders and glass pipe observed under analyzer and polarizer uncrossed. c) Detail of cylinder around the black line under crossed P/A. d) Scheme of layers in a cylinder. The dotted line represents the average molecular projections.*

I thank K.Djellab* for SEM imagery

*Plateforme Microscopie Electronique, Université de Picardie, 33 rue St Leu, 80039 Amiens

Correspondence: hugues.vasseur@u-picardie.fr